\documentclass[
prc,%
10pt,%
final,%
notitlepage,%
oneside,%
onecolumn,%
nobibnotes,%
nofootinbib,% 
superscriptaddress,%
floatfix,%
showkeys,%
showpacs]%
{revtex4-1}

\usepackage{color} 
\usepackage{amsfonts} 
\usepackage{amsbsy} 
\usepackage{mathrsfs}
\usepackage{graphicx}
\usepackage{comment}
\usepackage{pstricks,epsfig}
\usepackage{mathrsfs,slashed}
\usepackage{amsmath,amssymb}

\usepackage[most]{tcolorbox}

\def\lsim{\mathrel{\rlap{ \lower4pt\hbox{\hskip-3pt$\sim$}}
    \raise1pt\hbox{$<$}}} %less than approx. symbol
\def\gsim{\mathrel{\rlap{ \lower4pt\hbox{\hskip-3pt$\sim$}}
    \raise1pt\hbox{$>$}}} %greater than or approx. symbol

\preprint{draft \today}

\begin{document}
% ============================================================

\title{Quark cluster expansion model for interpreting finite-T lattice QCD thermodynamics} 
\author{David Blaschke}\thanks{e-mail: blaschke@ift.uni.wroc.pl}
\affiliation{Institute of Theoretical Physics, University of Wroclaw,
  50-204 Wroclaw, Poland} 
\affiliation{Bogoliubov Laboratory of
  Theoretical Physics, JINR Dubna, 141980 Dubna, Russia}
\affiliation{National Research Nuclear University (Moscow Engineering Physics Institute),
  115409 Moscow, Russia}
\author{Kirill~A.~Devyatyarov}\thanks{e-mail: dka005@campus.mephi.ru}
\affiliation{Bogoliubov Laboratory of
  Theoretical Physics, JINR Dubna, 141980 Dubna, Russia}
\affiliation{National Research Nuclear University (Moscow Engineering Physics Institute),
  115409 Moscow, Russia}
\author{Olaf Kaczmarek }\thanks{e-mail: okacz@physik.uni-bielefeld.de}
\affiliation{Fakult{\"a}t f{\"u}r Physik, Universit{\"a}t Bielefeld, Germany}
\affiliation{Key Laboratory of Quark \& Lepton Physics (MOE) and Institute of Particle Physics, Central China Normal University, Wuhan 430079, China}

\begin{abstract}
In this work, we present a unified approach to the thermodynamics of hadron--quark--gluon matter at finite temperatures 
on the basis of a quark cluster expansion in the form of a generalized Beth--Uhlenbeck approach with a generic ansatz for the hadronic phase shifts
that fulfills the Levinson theorem.
The change in the composition of the system from a hadron resonance gas to a quark--gluon plasma takes place in the narrow temperature interval of
150--190 MeV, where the Mott dissociation of hadrons is triggered by the dropping quark mass as a result of the restoration of chiral symmetry.
The deconfinement of quark and gluon degrees of freedom is regulated by the Polyakov loop variable that signals the breaking of the $Z(3)$ center 
symmetry of the color $SU(3)$ group of QCD. 
We suggest a Polyakov-loop quark--gluon plasma model with $\mathcal{O}(\alpha_s)$ virial correction and solve the stationarity condition of the thermodynamic potential (gap equation) for the Polyakov loop.
The resulting pressure is in excellent agreement with lattice QCD simulations up to high temperatures.

\end{abstract}
\pacs{11.10.Wx,12.38.Gc, 12.38.Mh, 25.75.Nq} 
\keywords{Polyakov quark--gluon plasma; hadron resonance gas; Beth--Uhlenbeck approach; lattice QCD thermodynamics}
%
% \date{\today}
%
\maketitle
% \today

\section{Introduction}
Recently, continuum extrapolated lattice QCD 
(LQCD) thermodynamics results for physical quark masses 
have become available  \cite{Borsanyi:2010bp,Bazavov:2011nk,Borsanyi:2013bia,Bazavov:2014pvz}.
It has now become a major goal to construct an effective low-energy QCD model that would 
reproduce these results in the finite temperature and low chemical potential domain to high accuracy.
Such a description could form a basis for extrapolations to the region of low temperatures and high 
baryochemical potentials where the sign problem still prevents LQCD obtaining benchmark solutions.

QCD Dyson--Schwinger equations \cite{Roberts:2000aa,Fischer:2018sdj} and functional renormalization 
group (FRG) methods  \cite{Tripolt:2013jra,Gao:2020qsj} are promising tools to investigate the nonperturbative 
aspects of the QCD phase diagram in the vicinity of the chiral and deconfinement transitions from first principles.
However, to date, they do not self-consistently account for the bound state formation.
Therefore, despite a satisfactory description of the temperature dependence of the renormalized 
light chiral condensate  \cite{Gao:2020qsj}, a question arises regarding the relationship to the chiral 
perturbation theory limit, where the onset of chiral condensate melting is solely due to light pseudoscalar 
meson excitations in the medium \cite{Gerber:1988tt}.   
The account for bound states in the FRG approach to low-energy QCD is still under development 
\cite{Alkofer:2018guy} and not yet applicable to address the role of hadronic bound states for chiral 
condensate melting.

Complementary to these first-principle approaches is the Polyakov--quark--meson (PQM) model 
(see, e.g., \cite{Herbst:2013ufa} and references therein), which is suitable to study the interrelation of 
chiral and deconfinement transitions. 
Both transitions are closely correlated, but in comparison with LQCD, the chiral restoration temperature 
result is too high. 
It could be suspected that the inclusion of further mesonic and baryonic degrees of freedom could improve 
the situation. Furthermore, the dynamic quark self-energy effects due to dressing by the meson cloud should be taken 
into account. In the PQM model, mesons are not composites but elementary degrees of freedom.

In order to properly account for the composite nature of mesons and baryons, as well as their Mott 
dissociation in a hot and dense medium, a Beth--Uhlenbeck approach can be employed
\cite{Hufner:1994ma,Zhuang:1994dw,Wergieluk:2012gd,Yamazaki:2012ux,Blaschke:2013zaa,Yamazaki:2013yua,Torres-Rincon:2016ahl,Torres-Rincon:2017zbr,Dubinin:2016wvt,Blaschke:2016fdh,Blaschke:2016hzu}.
This approach has been limited to the application of low-lying mesons only and did not take into account perturbative 
corrections, meaning that a quantitative description of LQCD thermodynamics has not yet been possible.

To overcome these limitations, the generic behavior of the scattering phase shifts in the hadronic  channels has been 
constructed in the spirit of a cluster expansion model which reproduces the full hadron resonance gas at low temperatures 
and the quark--gluon plasma (QGP) with $\mathcal{O}(\alpha_s)$ virial corrections at high temperatures  \cite{Blaschke:2016fdh,Blaschke:2016hzu}. 
The model embodied the main consequence of chiral symmetry restoration in the quark sector: the
lowering of the thresholds for the two and three-quark scattering state
continuous spectrum, which triggers the transformation of hadronic bound states to resonances in the scattering continuum.
In the early version of this model, the gap equation for the Polyakov-loop was incomplete, and the model of the phase shifts
was a rather complicated one.

We suggest here a Polyakov-loop quark--gluon plasma model with $\mathcal{O}(\alpha_s)$ virial correction 
in order to obtain a satisfactory agreement with  lattice QCD simulations up to high temperatures
and solve the complete stationarity condition of the thermodynamic potential (gap equation) for the Polyakov loop.
The phase shift model employed in this work is simpler than in previous works and universally applicable for all hadronic species.
It is in accordance with the Levinson theorem and results in the vanishing of hadronic contributions to the thermodynamics at high temperatures.

%%%%%%%%%%%%%%%%%%%%%%%%%%%%%%%%%%%%%%%%%%
\section{Cluster Virial Expansion to Quark-Hadron Matter}

The main idea behind unifying the description of the quark--gluon plasma (QGP) and the hadron resonance gas (HRG) phase of low-energy QCD matter is 
the fact that hadrons are strong, nonperturbative correlations of quarks and gluons. 
In particular, mesons and baryons are bound states (clusters) of quarks  and should therefore emerge in a cluster expansion of interacting quark matter
as new, collective degrees of freedom.  

For the total  thermodynamic potential of the model, from which all other equations of state can be derived, we make the following ansatz:
\begin{eqnarray}
\Omega_{\rm total}(T;\phi) = \Omega_{\rm QGP}(T;\phi) + \Omega_{\rm MHRG}(T) ~,
\end{eqnarray}
where $\Omega_{\rm QGP}(T;\phi)=\Omega_{\rm PNJL}(T;\phi)+ \Omega_{\rm pert}(T;\phi)$ 
describes the thermodynamic potential of the quark and gluon degrees of freedom with a 
perturbative part $\Omega_{\rm pert}(T;\phi)$ and a nonperturbative mean field part $\Omega_{\rm PNJL}(T;\phi)=\Omega_{\rm Q}(T;\phi)+\mathcal{U}(T;\phi)$ 
that can be decomposed into the quark quasiparticle contribution $\Omega_{\rm Q}(\phi;T)$ and the gluon contribution that is approximated by a mean field potential $\mathcal{U}(T;\phi)$.
Note that all these contributions to the QGP thermodynamic potential are intertwined by the traced Polyakov loop $\phi$ as the order parameter for confinement.
The correlations beyond the mean field approximation which correspond to the hadronic bound states and their scattering state continuum are described by the 
Mott--HRG pressure $P_{\rm MHRG}(T)$. This is an HRG pressure that takes into account the dissociation of hadrons by the Mott effect, when their masses would exceed the mass of the corresponding continuum of unbound quark states. 
A detailed description and numerical evaluation of these contributions is given below.

\subsection{Beth--Uhlenbeck Model for HRG with Mott Dissociation}

For the MHRG part of the pressure of the model, we have $P_{\rm MHRG}(T)= - \Omega_{\rm MHRG}(T)$,
\begin{eqnarray}
\label{eq:MHRG}
P_{\rm MHRG}(T) = \sum_{i = M,B}{P_i(T)}~,
\end{eqnarray}
where the sum extends over all mesonic (M) and baryonic (B) states from the particle data group (PDG), comprising an ideal mixture of hadronic bound and scattering states in the channel $i$ that are described by a Beth--Uhlenbeck formula. 
Then, the partial pressure of the hadron species $i$ is
\begin{equation}
\label{BU}
P_i(T)=\mp d_i\int_0^\infty\frac{dp~p^2}{2\pi^2}\int_0^\infty \frac{dM}{\pi}
	~T \ln \left(1\mp {\rm e}^{-\sqrt{p^2+M^2}/T}\right) \frac{d \delta_i(M;T)}{dM}~,
\end{equation}
where $d_i$ is the degeneracy factor. 
For the phase shift of the bound states of $N_i$ quarks in the hadron $i$, we adopt
the simple model that is in accordance with the Levinson theorem:
\begin{equation}
\label{phase}
\delta_i(M;T)=\pi\left[\Theta(M-M_i) - \Theta(M-M_{{\rm thr},i}(T))\right] \Theta(M_{{\rm thr},i}(T)-M_i).
\end{equation}

Inserting (\ref{phase}) into (\ref{BU}) results in 

\nointerlineskip
\vspace{3pt}
\begin{equation}
\label{BU-Mott}
P_i(T)=\mp d_i\int_0^\infty\frac{dp~p^2}{2\pi^2}
	~T \left[\ln \left(1\mp {\rm e}^{-\sqrt{p^2+M_i^2}/T}\right)
	- \ln \left(1\mp {\rm e}^{-\sqrt{p^2+M_{{\rm thr},i}(T)^2}/T}\right)\right]~\Theta(M_{{\rm thr},i}(T)-M_i).
\end{equation}

The temperature dependent threshold mass of the two (or three) quark continuum for mesonic (baryonic)
bound state channels $i$ is
\begin{equation}
\label{M_thr}
M_{{\rm thr},i}(T) = \sqrt{2}\left[
(N_i - N_s)m(T) + N_s m_s(T)
\right]~,
\end{equation}
where $N_s=0,1, ..., N_i$ is the number of strange quarks in hadron $i$.
The factor $\sqrt{2}$ originates from quark confinement in the following way.
In the confining vacuum, the quarks are not simple plane waves with an arbitrarily long wavelength,
but due to the presence of bag-like boundary conditions, their wavelength shall not exceed 
a certain length scale. Therefore, a minimal quark momentum applies to the quark 
dispersion relations $E_{q,{\rm min}}(T)=\sqrt{m_q^2(T)+p^2_{q,{\rm min}}}$, which for  the 
choice $p_{q,{\rm min}}=m_q(T)$ results in  $E_{q,{\rm min}}(T)=\sqrt{2}m_q(T)$.
For details, see \cite{Dubinin:2013yga}.
The chiral condensate is defined as
\begin{equation}
\label{cond}
\langle \bar{\psi} \psi \rangle_{q,T}=-\frac{\partial \Omega(T)}{\partial m_q}~,~~q=u,d,s,
\end{equation}
where $m_l$ ($m_s$) is the current-quark mass in the light (strange) quark sector, $l=u,d$.
It is an order parameter for the dynamical breaking of the chiral symmetry that is reflected in the 
corresponding temperature dependence of the dynamical quark masses $m_q(T)$.  

\begin{figure}[htb]
\centering
\includegraphics[width=0.73\textwidth]{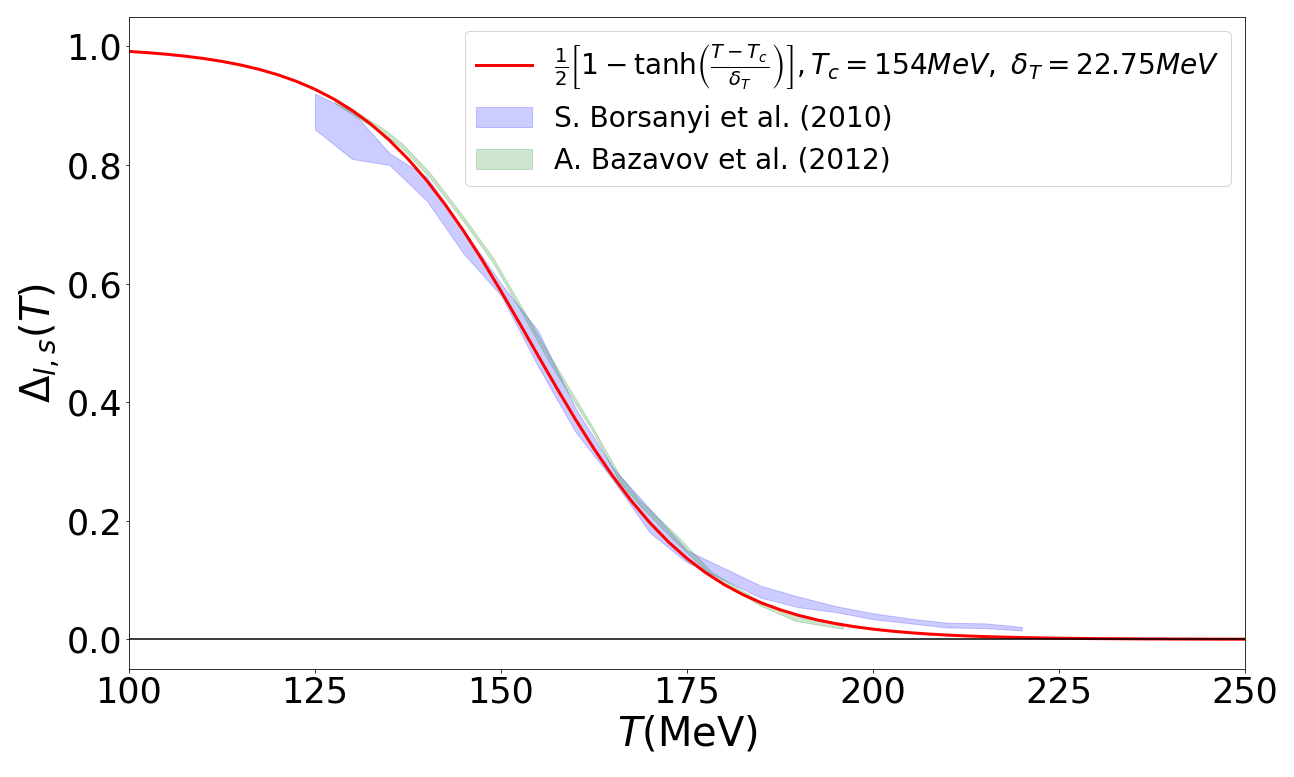}
\caption{Comparison of the fit (\ref{delta}) for the temperature dependence of the chiral condensate $\Delta_{l,s}(T)$
and the {LQCD} data for it from the Wuppertal--Budapest Collaboration \cite{Borsanyi:2010bp} and the hotQCD Collaboration
\cite{Bazavov:2011nk}.}
\label{fig1}
\end{figure}

In our present model, we do not treat the dynamical quark mass as an order parameter that should 
follow from  the solution of an equation of motion (gap equation) that minimizes the thermodynamic 
potential, as in the case of the Polyakov-loop variable $\phi$, but we use the quantity 
$\Delta_{l,s}(T)$ from simulations of 2 + 1 flavor lattice QCD as an input. 
This quantity has been introduced in \cite{Cheng:2007jq} with the definition
\begin{equation}
\label{Delta-ls}
\Delta_{l,s}(T)=\frac{\langle \bar{\psi} \psi \rangle_{l,T}-(m_l/m_s)\langle \bar{\psi} \psi \rangle_{s,T}}
{\langle \bar{\psi} \psi \rangle_{l,0}-(m_l/m_s)\langle \bar{\psi} \psi \rangle_{s,0}}~,
\end{equation}
and was used later on, e.g., in  \cite{Borsanyi:2010bp,Bazavov:2011nk}.
Further, we assume the following for the temperature-dependent light quark mass,
\begin{equation}
\label{m_light}
m(T)=m(0)\Delta_{l,s}(T)+m_l~,
\end{equation}
where $m_l=5.5$ MeV is the current-quark mass,
and for the strange quark mass, we adopt
\begin{equation}
\label{m_strange}
m_s(T) = m(T) + m_s - m_l=m(0)\Delta_{l,s}(T)+m_s~,
\end{equation}
with $m_s = 100$ MeV.
The LQCD result for the temperature dependence of the chiral condensate  
\cite{Borsanyi:2010bp,Bazavov:2011nk}  can be fitted by
\begin{equation}
\Delta_{l,s}(T)=\frac{1}{2}\left[1 - \tanh \left(\frac{T-T_c}{\delta_T} \right)\right]~,
\label{delta}
\end{equation}
where $T_c=154$ MeV is the common pseudocritical temperature of the chiral restoration
transition of both LQCD Collaborations and $\delta_T=26$ MeV is its width for the data from 
\cite{Borsanyi:2010bp}, while 
$\delta_T=22.7$ MeV for those from \cite{Bazavov:2011nk}, see Figure~\ref{fig1}.
For our present applications modeling QCD thermodynamics, we use the fit of the 
chiral condensate (\ref{delta}), but with the modern value of $T_c=156.5\pm1.5$ MeV \cite{Bazavov:2018mes}.
We have checked that the results for the total pressure of our model are practically inert
against a changing value of $\delta_T$ within the above range of variation. 
Inserting (\ref{m_light}) and (\ref{m_strange}) into (\ref{M_thr}), we get 
\begin{eqnarray}
M_{{\rm thr},i}(T) &=& \sqrt{2}\left[N_i m(T) + N_s (m_s(T)-m(T)) \right]\nonumber\\
&=& \sqrt{2} \left[m_s N_s + m_l (N_i - N_s) + m(0) N_i \Delta_{l,s}(T)\right]~,
\end{eqnarray}
and using (\ref{m_light}) results in
\begin{equation}
M_{{\rm thr},i}(T) =  \sqrt{2} \left\{ m_s N_s + m_l (N_i - N_s) + m(0) N_i\left[\frac{1}{2} - \frac{1}{2} \tanh \left(\frac{T-T_c}{\delta_T} \right)\right]\right\}~.
\end{equation}

\newpage

\begin{figure}[htb]
\centering
\includegraphics[width=0.73\textwidth]{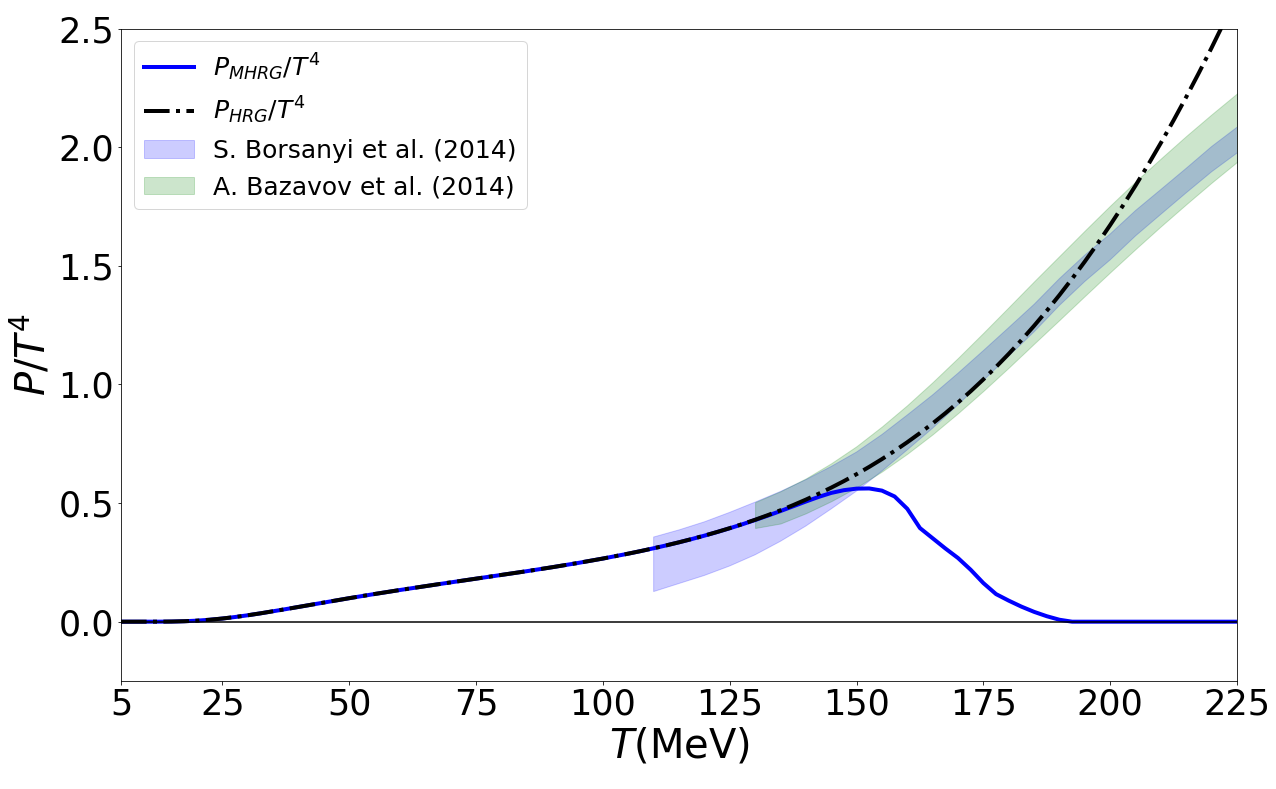}
\caption{Pressure as a function of temperature for the HRG model with stable hadrons (red line) and for the MHRG model with 
Mott dissociation of hadrons according to the simple phase shift model (\ref{phase}) employed in the present work.
These results are compared to the LQCD data from the HotQCD Collaboration \cite{Bazavov:2014pvz} (green band) and the
Wuppertal--Budapest Collaboration \cite{Borsanyi:2013bia} (blue band).
\label{fig:MHRG-HRG}
} 
\end{figure}

In Fig.~\ref{fig:MHRG-HRG} we show the pressure (\ref{eq:MHRG}) as a function of temperature for the hadron resonance gas (HRG) model with stable hadrons (red line) and for the HRG with Mott dissociation of hadrons (MHRG) according to the simple phase shift model (\ref{phase}) employed in the present work.
These results are compared to the LQCD data from the HotQCD Collaboration \cite{Bazavov:2014pvz} (green band) and the
Wuppertal--Budapest Collaboration \cite{Borsanyi:2013bia} (blue band). 
We want to point out that due to the Mott dissociation effect hadrons completely vanish from the system at $T\approx 190$ MeV 
while the pressure of the ideal HRG model is misleadingly still in perfect agreement with LQCD data at this temperature!
 
\subsection{Polyakov-Loop Improved Nambu--Jona-Lasinio (PNJL) Model}

The underlying quark and gluon thermodynamics are divided into a perturbative contribution 
$\Omega_{\rm pert}(T)$ which is 
treated as virial correction in two-loop order following~\cite{Turko:2011gw} and a nonperturbative   
part described within a {PNJL} 
model in the form
\begin{equation}
\label{P_PNJL}
P_{\rm PNJL}(T;\phi) = P_Q(T;\phi) + \mathcal{U}(T;\phi) ~,
\end{equation}
where the quark quasiparticle contribution is given by
\begin{equation}
P_Q(T;\phi) = 4N_c \sum_{q=u,d,s} \int \frac{dp~p^2}{2\pi^2} \frac{T}{3}\ln\left[1+3\phi(1+Y_q)Y_q+Y_q^3\right],~Y_q=e^{-\sqrt{p^2+m_q^2(T)}/T},
\end{equation}
and the Polyakov-loop potential $\mathcal{U}(T;\phi)$ takes into account the nonperturbative gluon background in a 
meanfield approximation using the polynomial fit of \cite{Ratti:2005jh}
\begin{equation}
\label{eq:U}
\mathcal{U}(T;\phi)=\frac{b_2(T)}{2}\phi^2 + \frac{b_3}{3}\phi^3 - \frac{b_4}{4}\phi^4~,
\end{equation}
where the temperature-dependent coefficient $b_2(T)$ is given by 
\begin{equation}
b_2(T)=a_0 + a_1 \left(\frac{T_0}{T}\right)
+ a_2 \left(\frac{T_0}{T}\right)^2
+ a_3 \left(\frac{T_0}{T}\right)^3~,
\end{equation}
and the coefficients are given in Table~\ref{tab1}.
\newline
\begin{table}[htb]
\caption{Set of values for the Polyakov-loop potential $\mathcal{U}(T;\phi)$ \cite{Ratti:2005jh}.
\label{tab1}
}
\setlength{\tabcolsep}{7.95mm}
\begin{tabular}{cccccc}
\toprule
\boldmath{$a_0$}	& \boldmath{$a_1$}	& \boldmath{$a_2$}	& \boldmath{$a_3$}	& \boldmath{$b_3$}	& \boldmath{$b_4$}\\
\hline
6.75	& $-$1.95	    & 2.625		& $-$7.44		& 0.75		& 7.5	\\
\hline
\end{tabular}
\end{table}

\subsection{Perturbative Contribution}

\begin{figure}[htb]
\centering
\includegraphics[width=0.3\textwidth]{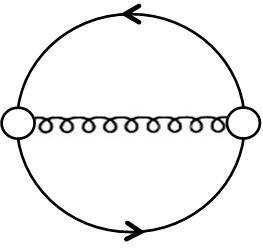}
\caption{Two-loop diagram for the contribution of the one-gluon exchange interaction to the thermodynamic potential of quark matter.
\label{fig:2loop}}
\end{figure}

It is well known that the lattice QCD thermodynamics at high temperatures of $T\sim 1$ GeV follow a Stefan--Boltzmann like behavior $\propto T^4$, but with a $15-20\%$ reduction of the effective number of degrees of freedom.
It has been observed, e.g., in \cite{Turko:2011gw}, that this deviation can be described by the virial correction to the pressure due to the 
quark--gluon scattering at $\mathcal{O}(\alpha_s)$ shown in Figure~\ref{fig:2loop}. 
Here, we modify the standard expression \cite{Kapusta:1989tk} of the~form
\begin{equation}
\Omega_{\rm pert}(T;\phi) = -\frac{8}{\pi} \alpha_s T^4
\left[
I(T;\phi) + \frac{3}{\pi^2} (I(T;\phi))^2
\right]
\end{equation}
by introducing the modified integral 
\begin{equation}
I(T;\phi)=\int_{\Lambda/T}^{\infty}{\rm d}x ~xf_{\phi}(x),
\end{equation}
where the generalized Fermi distribution function of the PNJL model for the case of vanishing
quark chemical potential considered here is defined as
\begin{equation}
f_{\phi}(x)=[\phi(1 + 2 Y) Y + Y^3]/[1+3\phi(1+ Y)Y + Y^3],~Y=\exp(-x)
\end{equation}
and $\Lambda=m_l(T)$ is the momentum range below which nonperturbative
physics dominates and is accounted for by the dynamically generated quark mass. 
Here, we use a temperature-dependent, regularized running coupling \cite{Peshier:2003ah,Blaschke:2005jg,Shirkov:2002td}
\begin{eqnarray}
\label{eq16}
\alpha_s = \frac{g^2}{4\pi} =\frac{12\pi}{11N_c-2N_f}\left( \frac{1}{\ln(r^2/c^2)} - \frac{c^2}{r^2-c^2} \right) ,
\end{eqnarray}
where $r =3.2 T$, $c = 350$ MeV and $N_c=N_f=3$.

\section{Stationarity Condition for the Polyakov Loop}

The pressure follows from the thermodynamic potential under the condition of stationarity with regard to variations of the order parameters.
Since the chiral condensate is fixed by the fit (\ref{delta}) to the numerical result from lattice QCD, 
the Polyakov loop $\phi$ is the only free-order parameter in the system to be varied; this condition means
\begin{equation}
P_{\rm QGP}(T) = - \min_{\phi} \left\{\Omega_{\rm QGP}(T;\phi)\right\}~.
\end{equation}
This is realized by demanding
\begin{equation}
\label{eq:phi}
\frac{d\Omega_{\rm QGP}(T;\phi)}{d\phi}=
\frac{dU(T;\phi)}{d\phi} +
\frac{d\Omega_Q(T;\phi)}{d\phi} +
\frac{d\Omega_{\rm pert}(T;\phi)}{d\phi}=0~,
\end{equation}
where the separate contributions come from the variations of the Polyakov loop potential
\begin{equation}
\frac{dU(T;\phi)}{d\phi}=b_2(T)\phi + b_3\phi^2 - b_4\phi^3~,
\end{equation}
and the quark quasiparticle pressure
\begin{equation}
\frac{d\Omega_Q(T;\phi)}{d\phi}  = 4N_c \sum_{q=u,d,s} \int \frac{dp~p^2}{2\pi^2}
\frac{(1+Y_q)Y_q}{1+3\phi(1+Y_q)Y_q+Y_q^3}~,
\end{equation}
with $Y_q=\exp[-\sqrt{p^2+m_q^2(T)}/T]$, and the $\mathcal{O}(\alpha_s)$ quark loop contribution 
\begin{equation}
\label{eq:25}
\frac{d\Omega_{\rm pert}(T;\phi)}{d\phi} = -\frac{8}{\pi} \alpha_s T^4
\left[
\frac{dI(\phi,T)}{d\phi} + \frac{6}{\pi^2} I(\phi,T)\frac{dI(\phi,T)}{d\phi}
\right]~,
\end{equation}
where
\begin{equation}
\label{eq:27}
\frac{dI(T;\phi)}{d\phi}=\int_{\Lambda/T}^{\infty}{\rm d}x ~x\frac{df_{\phi}(x)}{d\phi},
\end{equation}
and
\begin{equation}
\frac{df_{\phi}(x)}{d\phi}=\frac{Y+2Y^2-2Y^4-Y^5}{\left(1+3\phi(1+Y)Y+Y^3\right)^2}=
\frac{(1+2Y)Y-(2+Y)Y^4}{\left(1+3\phi(1+Y)Y+Y^3\right)^2}~,~Y=\exp(-x)~.
\end{equation}

The equation resulting from the stationarity condition (\ref{eq:phi}) can be dubbed a ``gap equation'' for $\phi$ since it has a similar structure to the quark mass gap equation, known from Nambu--Jona-Lasinio models. 
In previous work \cite{Blaschke:2016fdh,Blaschke:2016hzu}, the contribution to this gap equation from the $\mathcal{O}(\alpha_s)$ quark loop diagram was omitted.
Since this perturbative contribution is calculated with the Polyakov-loop generalized quark distribution functions $f_\phi$, it has to be included to the generalized gap equation for the traced Polyakov loop.   
This has been done in the present paper for the first time.

Moreover, an infrared cutoff is placed at the loop integrals in the perturbative contribution which is set at the medium-dependent quark mass. Therefore, the full perturbative contribution in accordance with \cite{Kapusta:1989tk} is restored only at high temperatures, where $m(T) \to m_l$ and $\phi \approx 1$, while in the vicinity of the chiral and deconfinement transition, the effects of both the quark mass and Polyakov loop are taken into account.
The solution of this gap equation gives the temperature dependence of the traced Polyakov loop $\phi$.
This is discussed in the next section. 

%%%%%%%%%%%%%%%%%%%%%%%%%%%%%%%%%%%%%%%%%%
\section{Results}

\subsection{Polyakov Loop}

\begin{figure}[htb]

\includegraphics[width=0.7\textwidth]{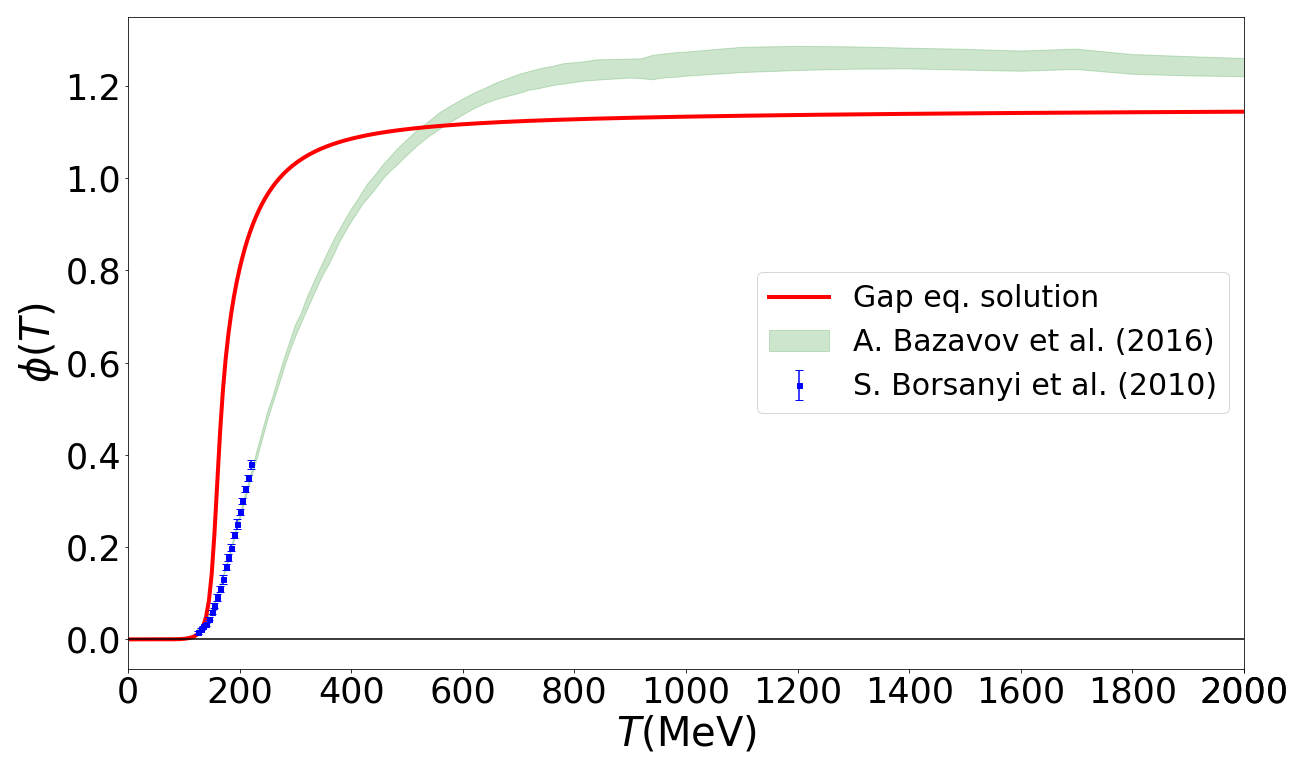}
\caption{The traced Polyakov loop $\phi$ from the solution of the stationarity condition (\ref{eq:phi}) on the thermodynamical potential as a function of temperature (magenta solid line)  compared with the lattice results for the renormalized Polyakov loop the {TU Munich QCD} (TUMQCD) 
Collaboration \cite{Bazavov:2016uvm} (green band) and the Wuppertal--Budapest Collaboration \cite{Borsanyi:2010bp} (blue symbols).}
\label{fig:phi}
\end{figure}

We performed the numerical solution of the gap Equation (\ref{eq:phi}) for the traced Polyakov loop as a function of the temperature which enters via the coefficient $b_2(T)$ of the Polyakov-loop potential (\ref{eq:U}) and the Boltzmann factors 
$Y_q$ and $Y$ of the distribution functions in the integrals (\ref{eq:25}) and (\ref{eq:27}). 
The result is shown in Figure~\ref{fig:phi} 
along with a comparison to the lattice QCD data for the renormalized Polyakov loop from the TUMQCD Collaboration~\cite{Bazavov:2016uvm} and the Wuppertal--Budapest Collaboration \cite{Borsanyi:2010bp}.

\begin{figure}[htb]
%\centering
\includegraphics[width=0.7\textwidth]{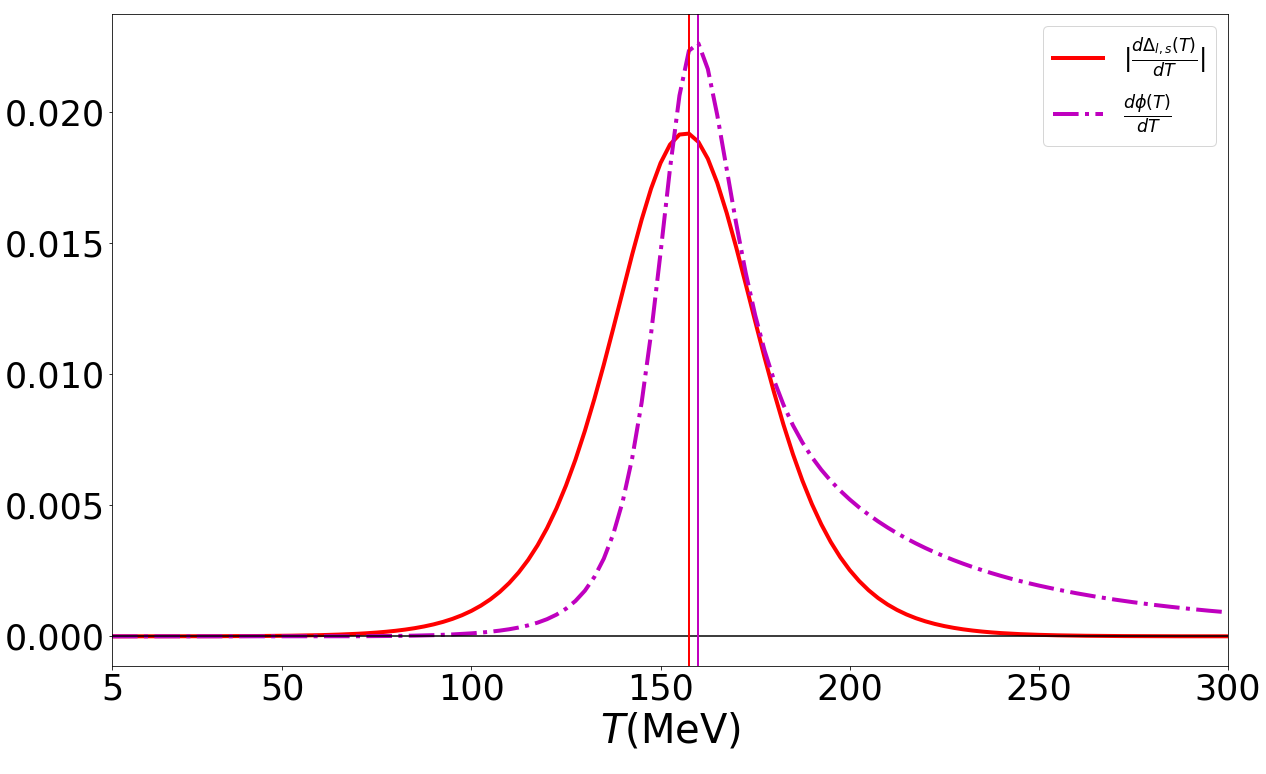}
\caption{The temperature derivatives of the chiral condensate (chiral susceptibility $d\Delta_{l,s}/dT$, red solid line) and of the Polyakov loop (Polyakov-loop susceptibility $d\phi/dT$, ) as functions of temperature. The vertical lines indicate their almost coincident peak positions at $T_\chi=156.5$ MeV and $T_\phi = 159.0$ MeV, respectively.
}
\label{fig:susc}
\end{figure}

In Figure~\ref{fig:susc}, we compare the Polyakov-loop susceptibility  $d\phi/dT$ with the chiral susceptibility $d\Delta_{l,s}/dT$
and obtain a strong synchronization effect of the chiral and Polyakov-loop crossover transitions. 
This is demonstrated by the almost coincident vertical lines indicating the peak positions of these transitions at $T_\chi=156.5$ MeV and $T_\phi = 159.0$ MeV, respectively.

\newpage
\subsection{Pressure}

\begin{figure}[htb]

\includegraphics[width=0.65\textwidth]{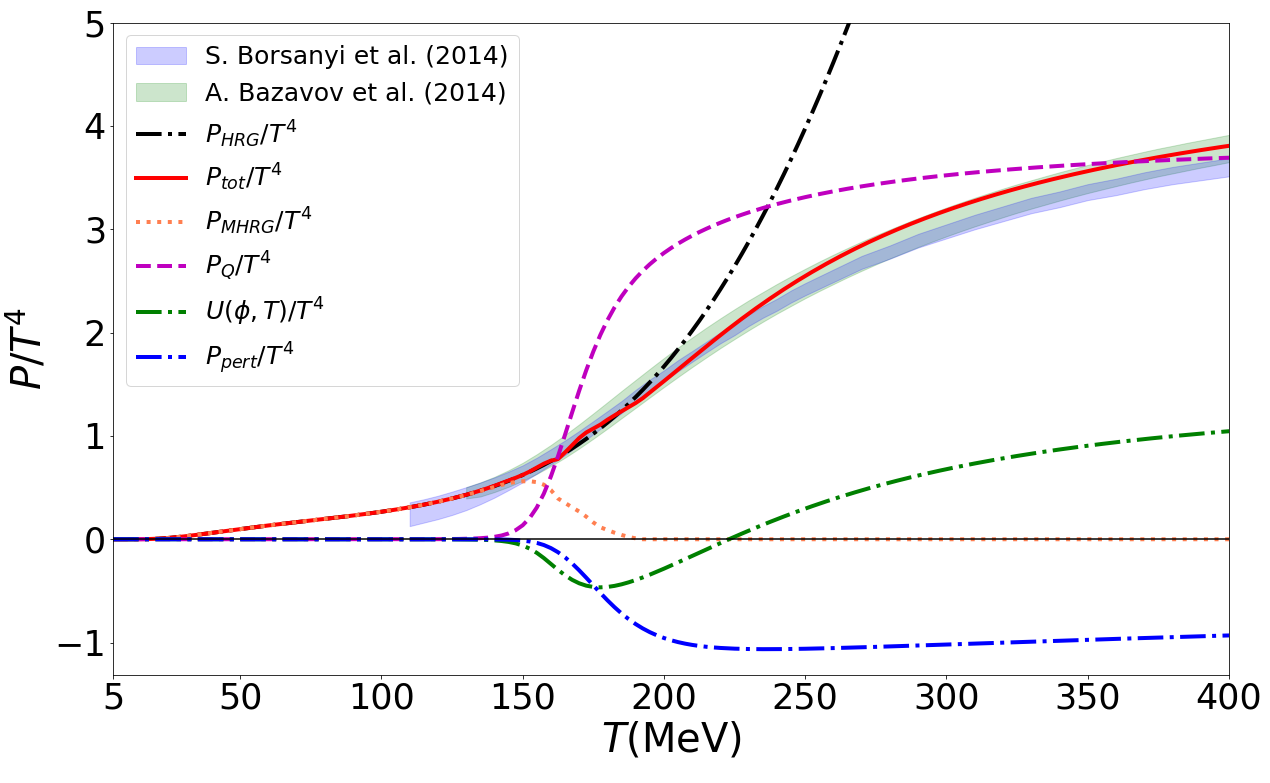}
\caption{{The} temperature dependence of the total scaled pressure (red solid line) and it's constituents: MHRG (coral dotted line), quark (dashed magenta line), Polyakov-loop potential $\mathcal{U}(T;\phi)$ (dash--dotted green line) and perturbative QCD contribution (dash-dotted blue line) compared to the lattice QCD data: HotQCD Collaboration \cite{Bazavov:2014pvz} (green band) and Wuppertal--Budapest Collaboration \cite{Borsanyi:2013bia} (blue band).
\label{fig:p-T}}
\end{figure}

\begin{figure}[htb]

\includegraphics[width=0.68\textwidth]{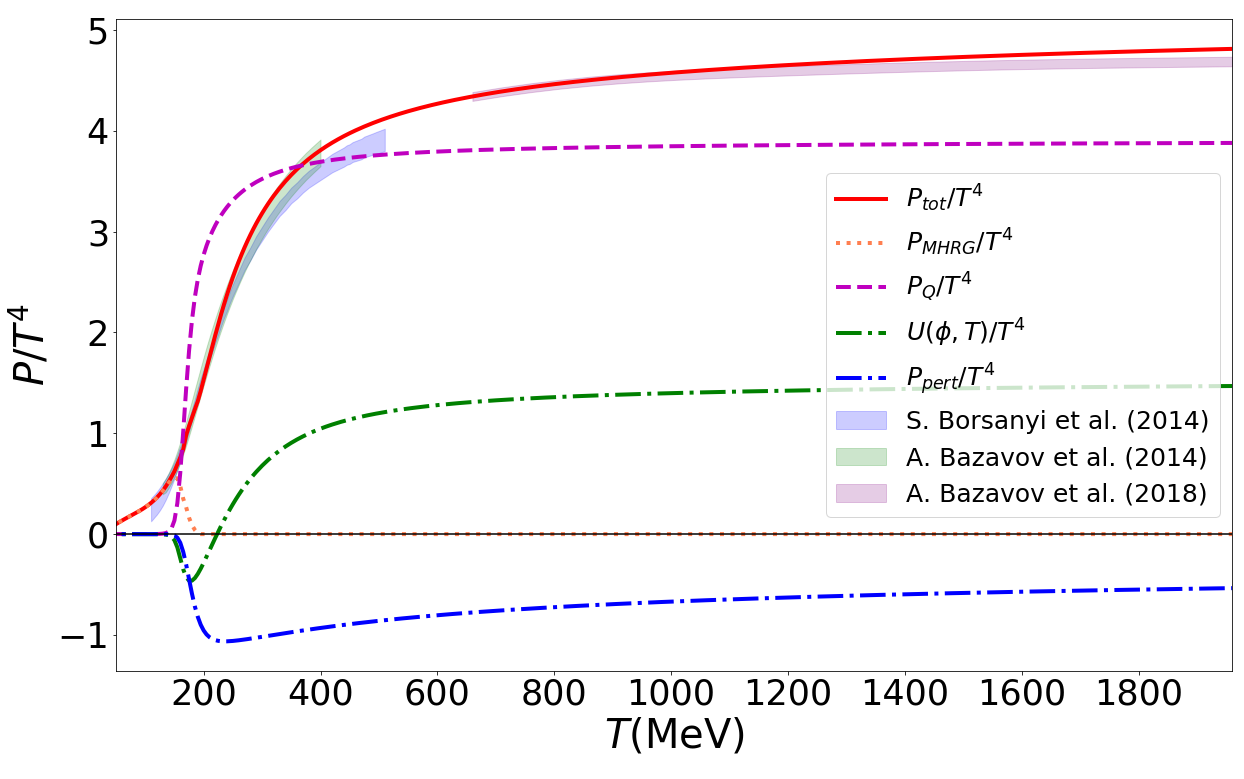}
\caption{{The} temperature dependence of the total scaled pressure (red solid line) and its constituents: MHRG (coral dotted line), quark (dashed magenta line), Polyakov-loop potential $U(\phi,T)$ (dash-dotted green line) and perturbative QCD contribution (dash-dotted blue line) compared to the lattice QCD data: HotQCD Collaboration \cite{Bazavov:2014pvz} (green band) and Wuppertal--Budapest Collaboration \cite{Borsanyi:2013bia} (blue band), and the high-temperature result 
\cite{Bazavov:2017dsy} (magenta band). 
\label{fig:p-T_high}}
\end{figure}

The main result of this work is a unified approach to the pressure of hadron--quark--gluon matter at finite temperatures that is in excellent agreement 
with lattice QCD thermodynamics (see Figure~\ref{fig:p-T}).
The nontrivial achievement of the presented approach is that the Mott dissociation of the hadrons described by the MHRG model pressure conspires 
with the quark--gluon pressure described by the Polyakov-loop quark--gluon model with $\mathcal{O}(\alpha_s)$ corrections in such a way that the 
resulting pressure as a function of temperature yields a smooth crossover behavior. 
By virtue of the Polyakov-loop-improved perturbative correction, the agreement with the lattice QCD thermodynamics extends to the high temperatures
of $T=1960$ MeV reported in \cite{Bazavov:2017dsy}; see Figure~\ref{fig:p-T_high}.  
  
 \vspace{-12pt}
 
 \newpage
 \subsection{Quark Number Susceptibilities}
 
In the present work, we did not consider the generalization of the approach to finite chemical potentials which would then allow us to evaluate the (generalized) 
susceptibilities as derivatives of pressure with respect to the corresponding chemical potential in appropriate orders. On that basis, ratios of susceptibilities could be formed as they indicate different aspects of the QCD transition between the limiting cases of a HRG and a QGP.
Here, we would like to discuss, as an insight into these extensions of the approach, one of the simplest susceptibility ratios, namely the dimensionless ratio of quark number density to quark number susceptibility:
\begin{equation}
\label{R12}
R_{12}(T)= \frac{n_q(T)}{\mu_q\, \chi_q(T)}\bigg{|}_{\mu_q=0}~,
\end{equation}
where $n_q(T)=\partial P(T,\mu_q)/\partial \mu_q|_{\mu_q=0}$ and $\chi_q(T)=\partial^2 P(T,\mu_q)/\partial \mu_q^2|_{\mu_q=0}$.
This ratio (\ref{R12}) has two well-known limits. At low temperatures, in the hadron resonance gas phase, it is given~by
\begin{equation}
\label{R12HRG}
R_{12}^{HRG}(T)= \frac{T}{3\mu_q} \tanh \left(\frac{3\mu_q}{T}\right)~,
\end{equation}
while in the QGP phase for massless quarks it approaches 
\begin{equation}
\label{R12QGP}
R_{12}^{QGP}(T)= \frac{1+(1/\pi^2)(\mu_q/T)^2}{1+(3/\pi^2)(\mu_q/T)^2}~.
\end{equation}
An evaluation of  (\ref{R12}) for the present model for the QCD pressure would require its extension to a finite $\mu_q$, which we will perform in a subsequent work.
In the present model, we used our knowledge of the composition as a function of temperature to define a proxy for (\ref{R12}) by interpolating between the two
known limits (\ref{R12HRG}) and (\ref{R12QGP}) with the partial pressure of the HRG, $x_{HRG}(T)=P_{\rm MHRG}(T)/P_{\rm tot}(T)$, as
\begin{equation}
\label{R12proxy}
R_{12}(T)= x_{HRG}(T) R_{12}^{HRG}(T) + [1-x_{HRG}(T)]R_{12}^{QGP}(T)~.
\end{equation}

\begin{figure}[htb]
\centering
\includegraphics[width=0.72\textwidth]{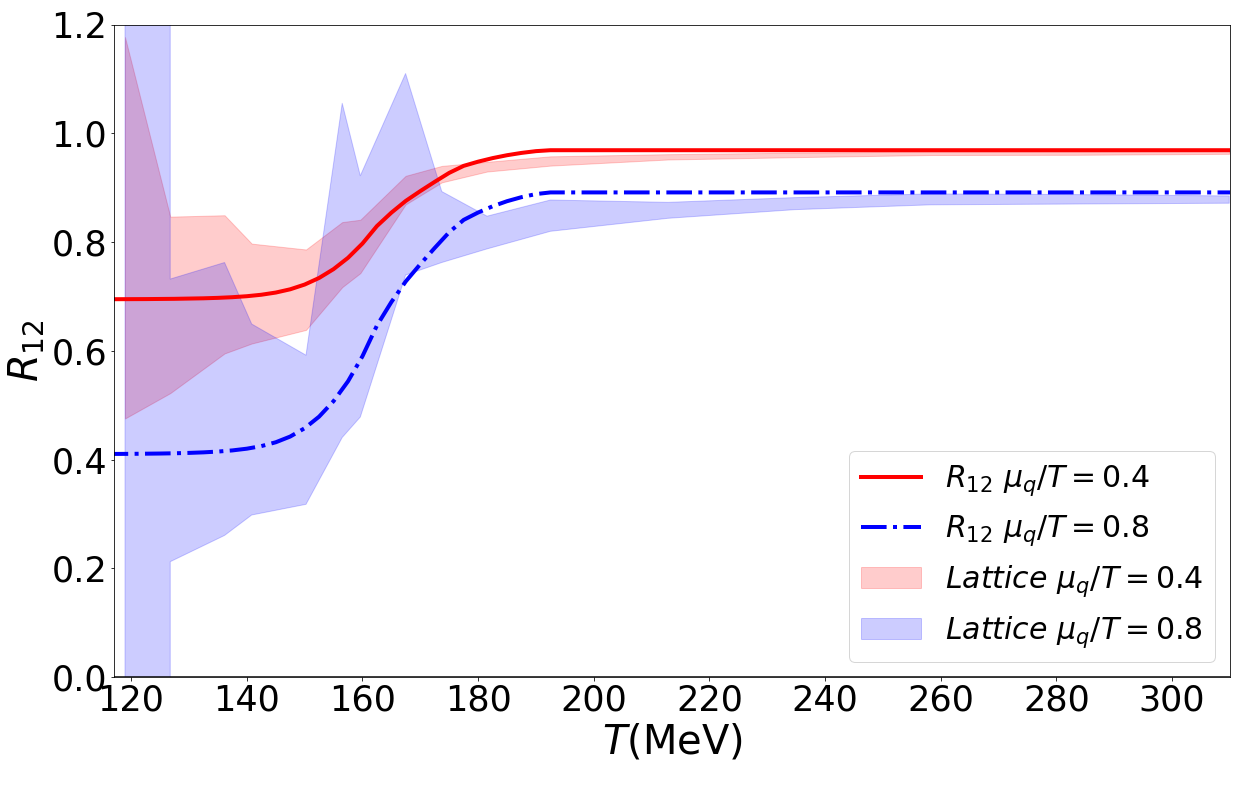}
\caption{The dimensionless ratio of quark number density to quark number susceptibility \mbox{$R_{12}(T)= n_q(T)/(\mu_q \chi_q(T))|_{\mu_q=0}$} as a function of temperature for $\mu_q/T = 0.4$ (red solid line) and $\mu_q/T = 0.8$ (blue dash-dotted line) compared to the lattice QCD data \cite{PhysRevD.71.054508} 
$\mu_q/T = 0.4$ (red band), $\mu_q/T = 0.8$ (blue band). For details, see text. 
\label{fig:R12}}
\end{figure}

The result is shown in Figure~\ref{fig:R12} for two values of $\mu_q/T$, for which lattice QCD results in the two-flavor case \cite{PhysRevD.71.054508}  are shown for a comparison.
The fact that the present approach reproduces the transition between HRG and QGP asymptotics well in the narrow range of temperatures $150~{\rm MeV}\lesssim T \lesssim 190~{\rm MeV}$ is a nontrivial result.
In previous effective approaches to describe the finite-temperature lattice QCD thermodynamics results based on a spectral broadening of the HRG states, the transition to the QGP asymptotics occurred at a much higher temperature 
 $250~{\rm MeV}\lesssim T \lesssim 400~{\rm MeV}$ \cite{Biro:2014sfa} or never \cite{Turko:2011gw,Turko:2013taa,Turko:2014jta}. 
 In the latter case, the QGP asymptotic behavior is mimicked by an appropriate number of unaffected low-lying hadronic degrees of freedom.
 
%%%%%%%%%%%%%%%%%%%%%%%%%%%%%%%%%%%%%%%%%%
\section{Discussion and Conclusions}
The main result of the present work is a unified approach to the thermodynamic potential of hadron--quark--gluon matter at finite temperatures that is in excellent agreement with lattice QCD thermodynamics on the temperature axis of the QCD phase diagram.
The key aspect to this approach is the quark cluster decomposition of the thermodynamic potential within the Beth--Uhlenbeck approach 
\cite{Bastian:2018wfl}, which allowed us to implement the effect of Mott dissociation to the hadron resonance gas phase of low-temperature/low-density QCD. 
The MHRG model description includes, in principle, the information about the spectral properties of all hadronic channels with their discrete and continuous part of the spectrum, encoded in the hadronic phase shifts.
Instead of solving the equations of motion with a coupled hierarchy of Schwinger--Dyson equations in the one, two and many-quark channels self-consistently 
(a formidable task of finite-temperature quantum field theory), we applied a schematic model for the in-medium phase shifts that was in accordance with the Levinson theorem and sufficiently general to be applicable for all multiquark cluster channels. 
This phase shift model requires only the knowledge of the vacuum mass spectrum which can come from the particle data group tables or from relativistic quark models and the medium dependence of the multi-quark continuum threshold.

The latter requires the knowledge of the quark mass (i.e., the chiral condensate) with its medium dependence as an order parameter of the chiral symmetry breaking and restoration. Since a quark mean field model of the (P)NJL  type is not sufficient as it lacks the back-reaction from the hadron resonance gas on the quark propagator properties, we employ here the chiral condensate measured in continuum-extrapolated, full lattice QCD with physical current quark masses as an 
input. This procedure restricts the applicability of the present model to small chemical potentials only, where lattice QCD data for the chiral condensate are available. In a further development of the model, a beyond-mean-field derivation of the quark self-energy will be given.
Furthermore, at the same level of approximation, the corresponding sunset-type diagrams for the $\Phi$ functional of the 2PI 
approach should be derived and evaluated. This would allow us to calculate the generalized polarization-loop integrals which determine the analytic properties of the multi-quark states. These can be equivalently encoded in the corresponding medium-dependent phase shifts of the generalized Beth--Uhlenbeck approach, as has been demonstrated in particular examples for pions, diquarks \cite{Blaschke:2013zaa,Blaschke:2014zsa} and nucleons \cite{Blaschke:2015sla} within the Polyakov-loop generalized NJL~model.   

Another important aspect of the present approach is that it leads to a relativistic density functional theory for QCD matter in the QCD phase diagram, with the known limits of the HRG and pQCD manifestly implemented. Such an approach allows us to predict the existence and location of critical endpoints in the QCD phase diagram, as has been demonstrated, e.g., in \cite{Bastian:2018mmc}, where a dependence on a free parameter could have---besides the critical endpoint of the liquid--gas transition in the nuclear matter phase---another endpoint for the deconfinement transition or none. This ``crossover all over'' case of the QCD phase diagram is impossible to address with two-phase approaches that use a Maxwell construction for the phase transition. 
Other models that are in use for the analysis of the critical behavior of QCD (see, e.g., \cite{Plumberg:2018fxo,Parotto:2018pwx}) impose this by assuming a so-called ``switch function'' between HRG and QGP phases. They are valuable tools but do not have predictive power.

%%%%%%%%%%%%%%%%%%%%%%%%%%%%%%%%%%%%%%%%%%
\vspace{6pt}

%%%%%%%%%%%%%%%%%%%%%%%%%%%%%%%%%%%%%%%%%%
\acknowledgments{D.B. acknowledges discussions with Konstantin Maslov, Krzysztof Redlich and Ludwik Turko about this work.}
%%%%%%%%%%%%%%%%%%%%%%%%%%%%%%%%%%%%%%
\newline
\newline
\textbf{Author contributions:} Conceptualization, D.B.; Data curation, K.A.D.; Formal analysis, D.B. and K.A.D.; Funding acquisition, D.B.; Investigation, K.A.D.; Software, K.A.D.; Supervision, D.B. and O.K.; Validation, O.K.; Writing original draft, D.B.; Writing review and editing, D.B. and O.K. All authors have read and agreed to the published version of the manuscript.
%%%%%%%%%%%%%%%%%%%%%%%%%%%%%%%%%%%%%%
\newline
\newline
\textbf{Funding:} The research of D.B. was supported by the Russian Fund for Basic research under grant number 18-02-40137 and from the National Research Nuclear University (MEPhI) in the framework of the Russian Academic Excellence Project under contract number 02.a03.21.0005.
%%%%%%%%%%%%%%%%%%%%%%%%%%%%%%%%%%%%%%
\newline
\newline
\textbf{Conflicts of interest:} The authors declare no conflict of interest.

\end{document}